# Data analytics approach to predict the hardness of the copper matrix composites

Somesh Kr. Bhattacharya, Ryoji Sahara
Research Center for Structural Materials, National Institute for Materials Science, 1-2-1 Sengen, Tsukuba, Ibaraki 305-0047, Japan

Dušan Božic, Jovana Ružić
Institute for Nuclear Sciences "Vinča", University of Belgrade, Mike Petrovića, Alasa 12-14, PO Box 522, 11001 Belgrade, Serbia

**Abstract**

Copper matrix composite materials have exhibited a high potential in applications where excellent conductivity and mechanical properties are required. In this study, the machine learning models have been applied to predict the hardness of the copper matrix composite materials produced via powder metallurgy technique. Two particular composites were considered in this work. From experiments, we extracted the independent variables (features) like the milling time (MT, Hours), dislocation density (DD, $m^{-2}$), average particle size (PS, μm), density ($\rho$, $gm/cm^3$) and yield stress ($\sigma$, MPa) while the Vickers Hardness (MPa) was used as the dependent variable. Feature selection was performed by calculation the Pearson correlation coefficient (PCC) between the independent and dependent variables. We employed six different machine learning regression models to predict the hardness for the two matrix composites.

**Keywords:** Copper Matrix Composites, Hardness, Machine Learning, Regression

## 1. Introduction

The excellent mechanical properties and electrical conductivity of copper matrix composites (CuMCs) and copper alloys possess [1, 2, 3] making them desirable materials in several industries viz. automotive, aerospace, military, nuclear, electronic. The main potential of these materials lies in reaching the favorable relation between improving the mechanical properties and preserving high conductivity. It is well known that the lower content of alloying elements in the copper matrix supports higher thermal and electrical conductivity. The most commonly used reinforcements [4, 5] for copper matrix are metals (Ti, Mg, Co, Ni, etc.) or ceramic particles SiC and $Al_2O_3$, while in recent years with particles such as $ZrB_2$, $TiO_2$, $TiB_2$, TiC, $B_4C$, etc. Since the properties of the CuMCs and its alloys strongly depends on the nature, amount and distribution of the reinforcements, the great attention is given to the selection of the manufacturing techniques for production of CuMCs and Cu alloys. Ingot and powder metallurgy are both used for production of the Cu based materials, where powder metallurgy is more suitable when in situ formation of the reinforcing particles is needed [6, 7, 8, 9, 10]. Although, the most recent study [11] of the copper matrix particulate-reinforced with $ZrB_2$ ceramics produced by ingot metallurgy show that as-cast $Cu-ZrB_2$ composites can reach the improvement in hardness up to 140 Vickers Hardness (HV) similar to the results obtained by powder metallurgy [12].

Investigation of the copper based materials attracts researches and engineers from different fields due their wide application and fast industry growth.

In powder metallurgy technique, the properties of alloys and metal matrix composites (MMCs) depend largely on the milling time. Thus, it is highly desirable to have a rapid and accurate prediction of the hardness via structure-property correlation of these MMCs. While physics-based models (e.g. density functional theory and phase field simulations) can promote understanding at a given length scale but they are often limited to low order model systems due to computational complexity and lack of input parameters to represent realistic higher-order systems. An efficient way to achieve is the data-driven methodology that involves applying statistical learning tools to analyze correlations between hardness and features of the MMCs. Machine learning (ML) approach can reduce the experimental cost and time while predicting target properties of materials [13, 14, 15, 16].

In the present study we made an attempt to apply ML approach to predict the hardness of the CuMCs. We employed six different regression models (random forest, gradient boost, near neighbor, support vector, kernel ridge and linear) to predict the hardness. The remaining paper is organized as follows. In section 2.1 and 2.2, we briefly describe the experimental work the machine learning model, respectively. We discuss our results in section 3 followed by conclusion.

**2. Methodology**

2.1. Experimental work
The Cu-ZrB2 alloy was produced using powder metallurgy technique, where Cu, Zr and B were used as starting powders. Mechanical alloying was performed in the attritor mill. The in situ formation of ZrB2 particles inside the Cu matrix was achieved during hot-pressing at 950°C. Morphological analyses of the mechanically alloyed (MA) powder mixtures were done by particle sizer and scanning electron microscopy (SEM). Microstructural characterization of the MA powder mixtures and hot-pressed samples were characterized by X-ray powder diffraction (XRD) and SEM. Detailed production procedure of Cu-ZrB2 composites and characterization methods applied have been described in previous studies [12, 17, 18].

2.2. Machine learning models

The primary requirement to build a statistical learning model for any material is to have a dataset containing the material descriptors or features, X. These descriptors represent the fundamental material properties. The basic task of the machine learning (ML) models if to map these features to a specific (target) property, Y (hardness in this case), that is, $Y = fX$. Thus, the two important elements of machine learning approach are the empirical model, $f$ and features, $X$. The ML model must be trained and cross validated using the training dataset which includes the measured targeted property. The trained model is then applied to an unseen dataset in order to predict he target property. From the experiments, we get the milling time (MT, Hours), dislocation density (DD, $m^{-2}$), average particle size (PS, $\mu m$) , density ($\rho$, $\frac{gm}{cm^3}$) and yield stress ($\sigma$ , MPa) as our descriptors, $X$. In this study we used two datasets two different MCs: (i) Cu-7% vol. $ZrB_2$ and (ii) Cu-2%vol. $ZrB_2$. As powder metallurgy is a time consuming process, in both the cases the datasets are small. As will be explain later, these small datasets are enough to understand the trend for these mechanically alloyed powders considered in the present study.

To predict the hardness of both the alloys, Cu-7%vol. $ZrB_2$ and Cu-2%vol. $ZrB_2$, we used different ML models: random forest (RF) regression [19], gradient boosting (GB) regression [20], support vector (SV) regression [21], k-Nearest neighbors (KNN) regression [22], linear regression (LR) [23] and kernel ridge (KR) regression [24] as implemented in the Python based open source data analytics toolkit, scikit-learn [25]. RF and GB regression models are ensemble learning methods where multiple decision trees are constructed. SV regression is considered a nonparametric technique as it relies on kernel functions. The linear regression models the relationship between the input and output variables using a linear predictor function and fits to minimize the residual sum of squares between observed data and predicted data. Kernel Ridge regression estimates the conditional expectation of a random variable to find a non-linear relationship between a pair of random variables. Using the kernel method, it simplifies the product of the inner products in a high dimensional space and learns a linear model in the implicit feature space induced by the kernel and the dataset. k-Nearest neighbors regression model uses a nonparametric method and outputs the average number of given data points, the k nearest neighbors.

Due to the availability of the small dataset, we performed Leave One Out (LOO) - cross validation (CV) [26]. The training of ML models with CV avoids the errors due the bias and variance. Finally, the hyperparameters for the ML models were optimized during the training process. For model performance we calculated the coefficient of determination, $R^2$ [27]. It is important to note that for both these CuMCs, we trained the ML models separately with their respective datasets.

## 3. Results and discussion

A strong influence of the milling parameters on the morphological and mechanical properties of alloys and MMCs has been reported in many studies. The duration of the milling process is essential in providing uniform distribution of the reinforcing particles in the metal matrix. During milling in the Attritor mill, the powder mixture is exposed to high energy collisions such as ball-particle-ball. Those collisions initiate changes in lattice parameters, shape and size as well as the hardness of the particles. Finding the suitable milling parameters for each alloy or composite material is a time consuming process.

First all the features were subjected to the correlation filter to remove those which are uncorrelated by calculating the Pearson correlation coefficient. The Pearson correlation is the measure of the linear correlation between the predictors, $X$, and target, $Y$. The Pearson correlation maps for both the CuMCs are shown in Figure 1. For both the MCs, we observed the yield stress and density to have the strongest correlation with hardness followed by dislocation density. The particle size was found to have the lowest correlation coefficient in case of Cu-7%.vol ZrB$_2$ composite. Importantly all the features were found to have a positive correlation coefficients for Cu-7%.vol ZrB$_2$ composite. For the case of Cu-2%.vol ZrB$_2$ composite, the features, milling time and particle size, were found to have negative correlation. While the milling time was found to have almost no correlation with hardness, the particle size was found to have a weak negative correlation. While we use all the features of Cu-7%.vol ZrB$_2$ composite for fitting the ML models, in case of Cu-2%.vol ZrB$_2$ composite we dropped the two feature milling time.

Table 1: The coefficient of determination ($R^2$) and mean absolute error (MAE) values obtained for the various ML methods applied to the Cu-7% vol. ZrB$_2$ composite are listed below.

| ML models | RF | KR | GB | SV | LR | KNN |
|---|---|---|---|---|---|---|
| $R^2$ | 0.92 | 0.92 | 0.88 | 0.84 | 0.83 | 0.79 |

Next we train the ML models using LOO-CV. We calculated the coefficient of determination ($R^2$) to evaluate the model performance. The coefficient of determination ($R^2$) which is calculated as

$$R^2 = 1 - \frac{\sum_{i=1}^{n}(y_i - \hat{y}_i)}{\sum_{i=1}^{n}(y_i - \bar{y})} \qquad (1)$$

where $y_i$ is the true value, $\hat{y}_i$ is the predicted value and $\bar{y}$ is the mean of $y_i$. The $R^2$ value lies between 0 and 1, with 1 signifying excellent fits.

In Table 1, we summarized the $R^2$ for Cu-7%.vol ZrB$_2$ composite. In this case, the random forest and kernel ridge regressor models exhibited the highest accuracy (92%) followed by gradient boosting regressor (88%) while the nearest neighbor regressor has the lowest accuracy of 79%. It is evident that all the models were able to achieve an accuracy of 80% or even higher. For the two best performing ML models, random forest and kernel ridge, we plotted the true and predicted values of hardness for the Cu-7 vol.% ZrB$_2$ composite as shown in Figure 2.

In Table 2, we tabulated the $R^2$ values obtained for the different ML models applied to Cu-2% vol. ZrB$_2$ composite. For gradient boosting we achieved an accuracy of 79% while for the support vector regressor and for kernel ridge regressor we obtained an accuracy of 74%. Overall, all the ML models have a lower accuracy in case of Cu-2vol.% ZrB$_2$ compared to Cu-7% vol. ZrB$_2$ composite. We think perhaps more data is necessary to make a better predictive model for the hardness of Cu-2% vol. ZrB$_2$ composite. In figure 3, we plotted the true and predicted values of hardness for the Cu-2% vol. ZrB$^2$ composite for the gradient boosting and random forest models.

Table 2: The coefficient of determination ($R^2$) and mean absolute error (MAE) values obtained for the various ML methods applied to the Cu-2% vol. ZrB$_2$ composite are listed below.

| ML models | SV | KR | LR | GB | RF | KNN |
|---|---|---|---|---|---|---|
| $R^2$ | 0.79 | 0.74 | 0.68 | 0.62 | 59 | 0.50 |

## 4. Conclusion

In summary, we have built a regression model to predict the hardness of CuMCs prepared by powdered milling method. For Cu-7 vol.% ZrB$_2$ composite we achieved an accuracy of 80% or higher. On the other hand, the ML models for Cu-2 vol.% ZrB$_2$ composite have a lower predictive accuracy. To improve the accuracy of the ML models, we think some more data points must be included in the training dataset. The same strategy can be extended to other matrix composites prepared from mechanical alloying method.

**Acknowledgments**
SKB and RS acknowledge the support from the project Council for Science, Technology and Innovation(CSTI), Cross-ministerial Strategic Innovation Promotion Program (SIP), "Materials Integration for revolutionary design system of structural materials" (Funding agency: JST). DB and JR acknowledge the financial support from the Ministry of Education and Science of the Republic of Serbia through the Project No 172005.

**References**
[1] K. U. Kainer, Metal matrix composites custom-made materials for automotive and aerospace engineering,, WILEY-VCH Verlag GmbH & Co. KGaA, Weinheim.

[2] S. R. Pogson, P. Fox, C. Sutcliffe, W. O'Neill, The production of copper parts using dmlr. rapid prototyping, PRB 9 (2003) 334343.

[3] M. Li, S. J. Zinkle, Comprehensive nuclear materials 4 (2012) 667–690.

[4] Y. Zhan, Y. Z. J. Zeng, Tribological properties of Al$_2$O$_3$/CuCrZr composites, Tribol Lett 20 (2005) 163170.

[5] M. Sobhani, A. Mirhabibi, H. Arabi, R. M. D. Brydson, Effects of in situ formation of TiB$_2$ particles on age hardening behavior of Cu1wt% Ti1wt%
TiB$_2$, Mater Sci Eng A 577 (2013) 1622.

[6] W. S. Miller, F. J. Humphreys, Strengthening mechanisms in particulate metal matrix composites, Scripta Metallurgica et Materialia 25 (1991) 33-38.

[7] C. Zou, Z. Chen, E. Guo, H. Kang, G. Fan, W. Wang, A nano-micro dual-scale particulate-reinforced copper matrix composite with high strength, high electrical conductivity and superior wear resistance, RSC Adv 8 (2018) 30777–30782.

[8] S. J. Dong, Y. Zhou, Y. Shi, B. Chang, Formation of a $TiB_2$-reinforced copper-based composite by mechanical alloying and hot pressing, Metallurgical and Materials Transactions A 33A.

[9] J. R. Groza, J. C. Gibeling, Principles of particle selection for dispersion strengthened copper, Materials Science and Engineering: A 171 (1-2) (1993) 115–125.

[10] H. Kimura, N.Muramatsu, K.Suzuki, Copper alloy and copper alloy manufacturing method, Patent Application Publication, US 0211346A1, United States.

[11] X. Fan, X. Huang, Q. Liu, H. Ding, H. Wang, C. Hao, The microstructures and properties of in-situ $ZrB_2$ reinforced cu matrix composites, Results in Physics 14 (2019) 102494.

[12] J. Ruzic, J. Stasic, S. Markovic, K. Raic, D. Bozic, Synthesis and characterization of Cu-$ZrB_2$ alloy produced by PM techniques, Science of Sintering 46, br.2 (2014) 217-224.

[13] D. Xue, D. Xue, R. Yuan, Y. Zhou, P. V. Balachandran, X. Ding, J. Sun, T. Lookman, An informatics approach to transformation temperatures of NiTi-based shape memory alloys, Acta Materialia 125 (2017) 532 – 541.

[14] D. Shin, Y. Yamamoto, M. Brady, S. Lee, J. Haynes, Modern data analytics approach to predict creep of high temperature alloys, Acta Materialia 168 (2019) 321-330.

[15] J. Wei, X. Chu, X.-Y. Sun, K. Xu, H.-X. Deng, J. Chen, Z. Wei, M. Lei, Machine learning in materials science, InfoMat 1 (3) (2019) 338-358.

[16] J. Schmidt, M. R. G. Marques, S. Botti, M. A. L. Marques, Recent advances and applications of machine learning in solid-state materials science, npj Computational Materials 5 (2019) 83.

[17] J. Ruzic, J. Stasic, V. Rajkovic, D. Bozic, Synthesis, microstructure and mechanical properties of $ZrB_2$ nano and microparticle reinforced copper matrix composite by in situ processings, Materials and Design 62 (2014) 409-415.

[18] J. Ruzic, J. Stasic, V. Rajkovic, K. Raic, D. Bozic, Microstructural and mechanical properties of Cu-7vol.%ZrB2 alloy produced by metallurgy processing techniques, Science and Engineering of Composite Materials 22 (2015) 665-671.

[19] M. R. Segal, Machine learning benchmarks and random forest regression, Biostatistics (2004) 1-14.

[20] L. Mason, J. Baxter, P. Bartlett, M. Frean, Boosting algorithms as gradient descent, In S.A. Solla and T.K. Leen and K. M¨uller (ed.) Advances in Neural Information Processing Systems 12. MIT Press. 12 (1999) 512-518.

[21] H. Drucker, C. C. Burges, L. Kaufman, A. J. Smola, V. N. Vapnik, Support vector regression machines, Advances in Neural Information Processing Systems 9, NIPS, MIT-Press 9 (1997) 155–161.

[22] V. Cherkassky, Y. Ma, Comparison of model selection for regression, Neural Comput. 15 (2003) 1691-1714.

[23] J. Neter, M. H. Kutner, C. J. Nachtsheim, W. Wasserman, Applied linear statistical models, Chicago: Irwin. 4.

[24] Y. Zhang, J. Duchi, M. Wainwright, Divide and conquer kernel ridge regression: A distributed algorithm with minimax optimal rate, Journal of Machine Learning Research 16 (2015) 3299-3340.

[25] F. Pedregosa, G. Varoquaux, A. Gramfort, V. Michel, B. Thirion, O. Grisel, M. Blondel, G. Louppe, P. Prettenhofer, R. Weiss, V. Dubourg, J. Vanderplas, A. Passos, D. Cournapeau, M. Brucher, M. Perror, E. Duchesnay, Scikit-learn: Machine learning in python 12 (2012) 2825–2830.

[26] C. Sammut, G. I. Webb, Leave-one-out cross-validation in encyclopedia of machine learning, Springer, Boston, MA.

[27] J. L. Devore, Probability and statistics for engineering and the sciences (8th ed.), Boston, MA: Cengage Learning (2011) 508-510.

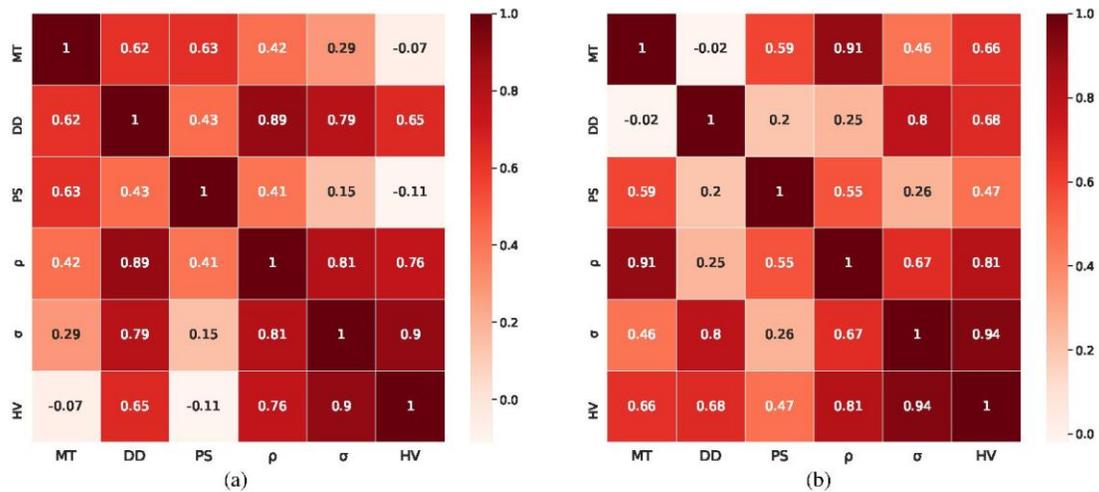

***Figure 1:*** *The Pearson correlation maps for the features and the target for (a) Cu-2%.volZrB$_2$, and (b) Cu-7%.vol ZrB$_2$. The color tone depicts the significance of the correlation.*

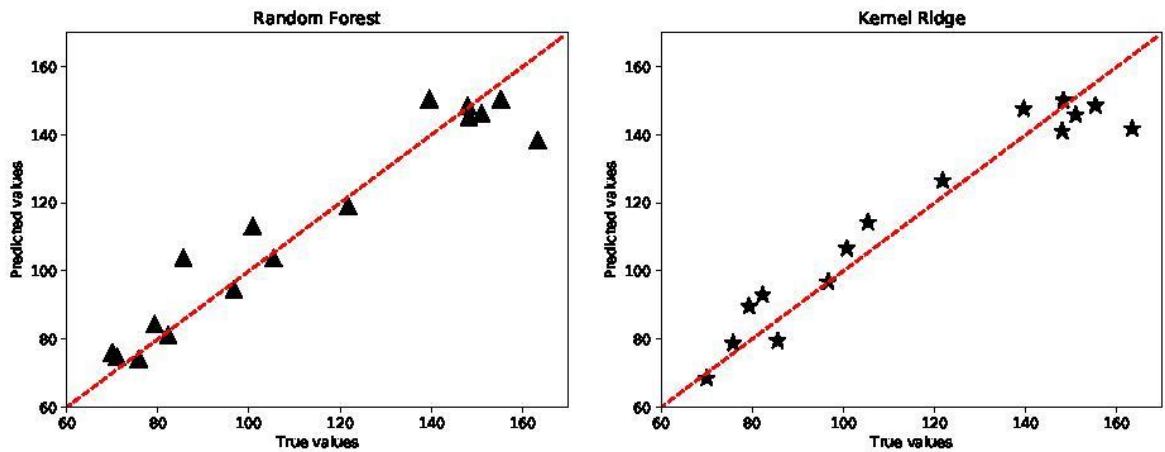

**Figure 2:** *The plots for the true values of Vickers hardness (experiment) and the predicted* values of hardness using the two best performing ML models for Cu-7%.vol ZrB2 composite are plotted. The broken red line depicts the case where the true and predicted values exactly match.

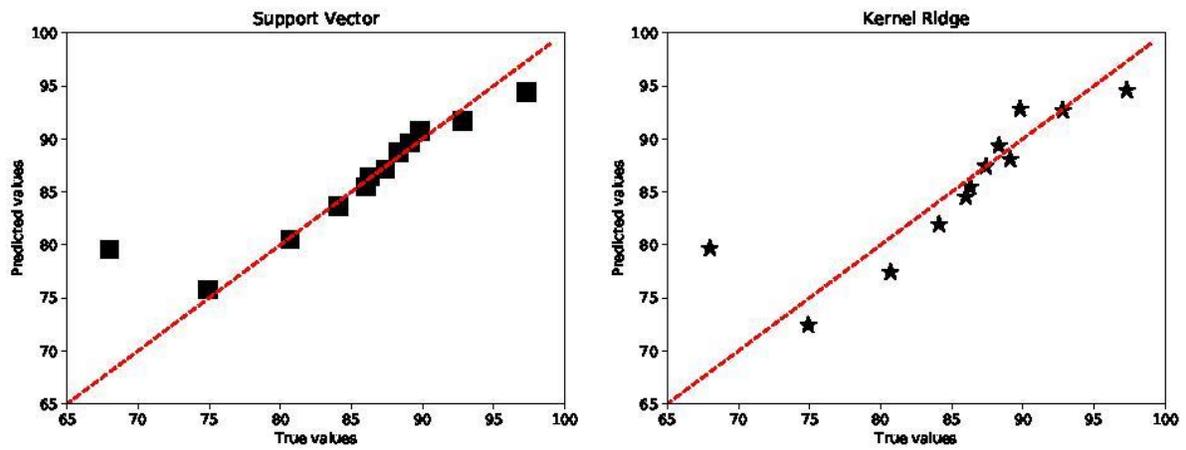

**Figure 3:** *The plots for the true values of Vickers hardness (experiment) and the predicted* values of hardness using the two best performing ML models for Cu-2%.vol ZrB$_2$ composite are plotted. The broken red line depicts the case where the true and predicted values exactly match.